\documentclass[iop]{emulateapj}
\usepackage{graphicx}
\usepackage{savesym}
\usepackage{listings}
\usepackage[intlimits]{amsmath}
\savesymbol{iint}
\usepackage{txfonts}
\usepackage{bm}
\restoresymbol{TXF}{iint}
\usepackage{amssymb}

\def\be{\begin{equation}}
\def\ee{\end{equation}}
\def\ba{\begin{eqnarray}}
\def\ea{\end{eqnarray}}
\def\go{\mathrel{\raise.3ex\hbox{$>$}\mkern-14mu
             \lower0.6ex\hbox{$\sim$}}}
\def\lo{\mathrel{\raise.3ex\hbox{$<$}\mkern-14mu
             \lower0.6ex\hbox{$\sim$}}}

\def\apj{ApJ}
\def\apjl{ApJL}
\def\apjs{ApJS}

\def\aap{A\&A }

\def\mnras{MNRAS}

\def\nat{Nature}

\begin{document}
\title{Timing Noise in Pulsars and Magnetars and the Magnetospheric Moment of Inertia}

\author{David Tsang and Konstantinos N. Gourgouliatos}
\affiliation{Physics Department McGill University, 3600 rue University, Montreal, QC, Canada, H3A 2T8}
\email{dtsang@physics.mcgill.ca, kostasg@physics.mcgill.ca}
\begin{abstract}
We examine timing noise in both magnetars and regular pulsars, and find that there exists a component of the timing noise ($\sigma_{\rm TN}$) with strong magnetic field dependence ($\sigma_{\rm TN} \sim B_o^{2} \Omega T^{3/2}$) above $B_o \sim 10^{12.5}$G. The dependence of the timing noise floor on the magnetic field is also reflected in the smallest observable glitch size. We find that magnetospheric torque variation cannot explain this component of timing noise. We calculate the moment of inertia of the magnetic field outside of a neutron star and show that this timing noise component may be due to variation of this moment of inertia, and could be evidence of rapid global magnetospheric variability.
\end{abstract}

\date{\today}
\maketitle

\section{Introduction}
The spin evolution of pulsars and magnetars is studied by long term monitoring of pulse arrival times, allowing the spin period, spin-down rate, orbital motion and astrometric variations of the neutron star (NS) to be inferred. The variation of these pulse arrival times from the best fit models of the timing evolution is referred to as timing noise. 

Timing noise was first observed in the Crab pulsar \citep{Boynton:1972} and has been identified as a ubiquitous property of pulsars \citep{Helfand:1980}. Recently much effort has been made to understand the stochastic timing noise in millisecond pulsars, as timing arrays of low-noise recycled pulsars will allow the detection of nanohertz gravitational waves \citep{Foster:1990}. Millisecond pulsars typically have relatively low inferred surface magnetic field ($B_o \sim10^7-10^8$G), due to age and accretion history.

Most isolated pulsars have spin-down inferred surface magnetic fields ($B_o$) in the range $B_o \sim 10^{11} - 10^{13}$G, while the highest known magnetic fields are possessed by magnetars ($B_o \sim 10^{13}-10^{15}$G). Magnetars include Anomalous X-ray Pulsars (AXPs) whose X-ray luminosities dwarf their spin down luminosities and most likely originate from internal magnetic field decay \citep{ThompsonDuncan:1996}; and Soft Gamma Repeaters (SGRs), hard X-ray transient sources that can undergo extreme outbursts. 

Mode-changing and nulling behaviors in pulsars have been linked to spin-down torque variations \citep{Lyne:2010, Kramer:2006}, implying that such behavior is indicative of variations of the open field line regions of the magnetosphere. Recent simultaneous X-ray and radio observations of PSR B0943$+$10 \citep{Hermsen:2013} have provided evidence that such variability is indicative of rapid {\it global} variations of the magnetosphere. 

In this Letter we study the timing noise for radio pulsars and magnetars, inferring a source of timing noise in high B-field pulsars and AXPs which correlates strongly with magnetic field. We examine physical models to explain the dependence of this timing noise on $B_o$, and identify this timing noise as evidence for global magnetospheric variability.

\section{Timing Noise Analysis}
Long term pulsar timing irregularities have enjoyed a long history of detailed analysis \citep[see e.g. ][]{Hobbs:2010, Shannon:2010, D'Alessandro:1995, Arzoumanian:1994, Cordes:1985}. Here we adopt the approach of modelling timing noise as a random walk process \citep{Boynton:1972, Groth:1975, Cordes:1980a}. We utilize the formalism of \citet{Cordes:1980a} in order to define the random walk strengths of various processes: $S_{\rm PN} = R\left<(\delta \phi)^2\right>$,  $S_{\rm FN} = R\left<(\delta \Omega)^2\right>$, and $S_{\rm SN} = R\left<(\delta \dot{\Omega})^2\right>$, for random walks in phase, frequency, and spin-down respectively, where $R$ is the occurrence rate of the random walk steps, and $\big< \cdot \big >$ indicates an ensemble average. The quantities $\delta \phi$, $\delta \Omega$, and $\delta \dot{\Omega}$ denote stochastic variations in phase, frequency, and spin-down rate, respectively.

These strengths can be estimated from timing parameters of a given pulsar,
\ba
S_{\rm PN} &\simeq& 2 C_{0,m}^2 \sigma_{\rm TN}^2 T^{-1},\\
S_{\rm FN} &\simeq& 12 C_{1,m}^2 \sigma_{\rm TN}^2 T^{-3},\label{fnstrength}\\
S_{\rm SN} &\simeq& 120 C_{2,m}^2 \sigma_{\rm TN}^2 T^{-5},
\ea
where, $T$ is the time span over which the observations were taken, $\sigma_{\rm TN}$ is the rms phase residual of the data for a given timing solution, and $C_{0,m}$, $C_{1,m}$ and $C_{2,m}$ are correction factors \citep{Cordes:1980a, Deeter:1984}  to compensate for random walk power removed by the $m$th-order polynomial fit when the residuals are determined. 

Previous analyses \citep{Cordes:1985, D'Alessandro:1995, Hobbs:2010} have shown that simple random walk processes cannot explain the totality of timing noise. We argue that if some random walk timing noise component becomes dominant as magnetic field increases from $\sim10^7$G to $\sim10^{15}$G this should result in a lower bound in the distribution of the random walk strength that increases with $B_o$. 

\begin{figure}
\includegraphics[width=\columnwidth]{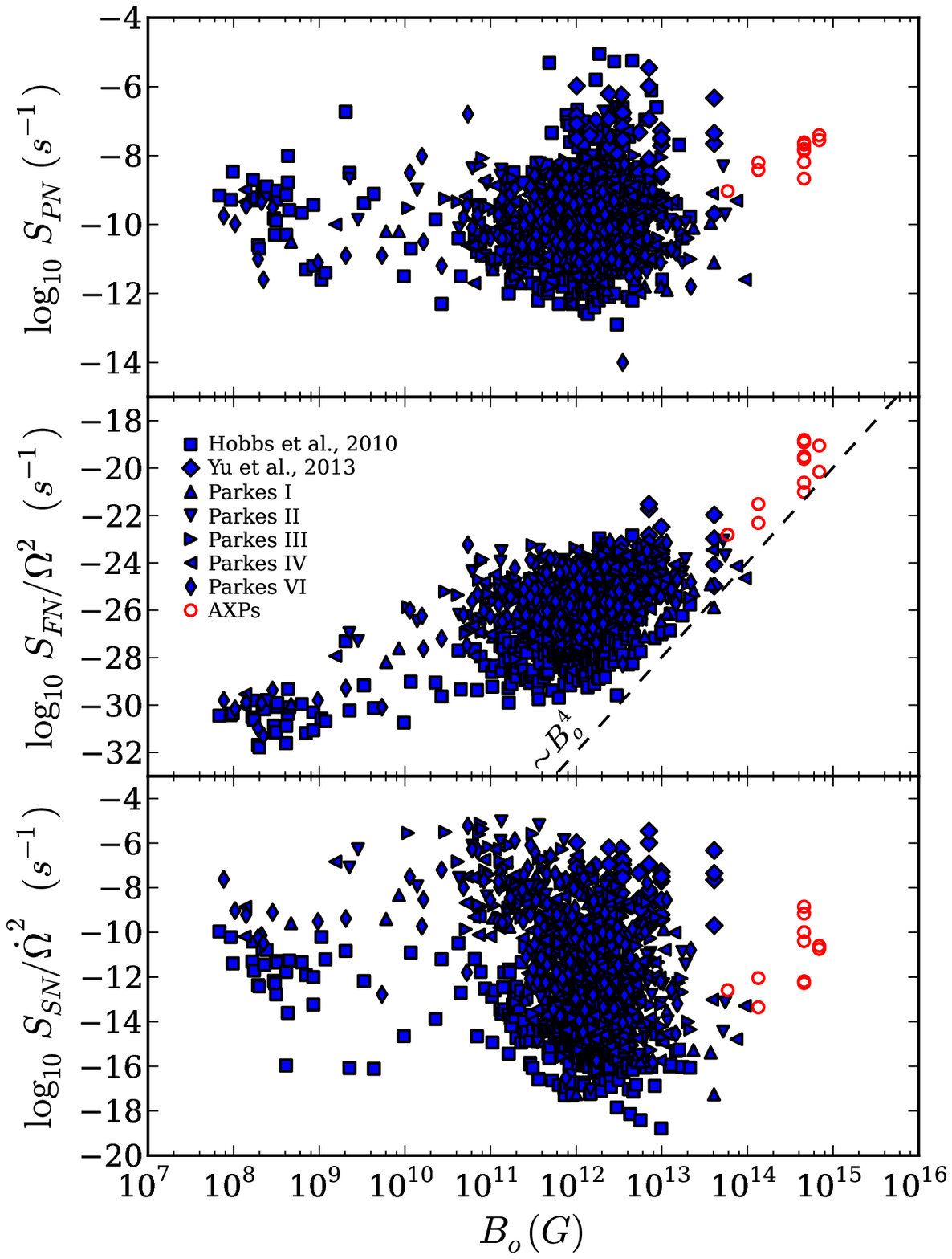}
\caption{Effective random walk strength versus $B_o$. The frequency noise (FN) and spin-down noise (SN) strengths have been normalized by $\Omega^2$ and $\dot{\Omega}^2$, respectively.  The blue points denote normal pulsars, with radio timing information taken from \citet{Hobbs:2010}, \citet{Yu:2013}, and the Parkes multi-beam survey \citep{ParkesI, ParkesII, ParkesIII, ParkesIV, ParkesVI}. The red circles label AXP timing epochs, with X-ray timing information taken from \citet{Dib:2008, Dib:2007}, and \citet{Gavriil:2002}. $S_{\rm FN}$ shows a strong trend with $B_o$. In particular the lower bound of the $S_{\rm FN}$ distribution appears to rise sharply for $B_o \go 10^{12.5}$ G as $(S_{\rm FN}/\Omega^2)_{min} \sim B_o^4$. \label{strengthplot}}
\end{figure}

In Figure \ref{strengthplot} we show the random walk strengths for various pulsars versus $B_o$ the surface dipole magnetic field strength. The strengths $S_{\rm FN}$ and $S_{\rm SN}$ have been normalized by $\Omega^2$ and $\dot{\Omega}^2$, respectively, to be comparable across pulsars with differing timing profiles. Here, we utilize published timing data from Jodrell Bank Observatory \citep{Hobbs:2010}, and the Parkes 64m radio telescope \citep{Yu:2013, ParkesI, ParkesII, ParkesIII, ParkesIV, ParkesVI}. We also include the X-ray timing results from the well-timed AXPs: 1E 1841$-$045 \citep{Dib:2008}, RXS J170849.0$-$400910 \citep{Dib:2008}, 4U 0142$+$61\citep{Dib:2007}, and 1E 2259.1$+$586 \citep{Gavriil:2002}. We do not include the timing for AXP 1E 1547.0$-$5408 \citep{Dib:2012}, as the timing observations have only been taken post-outburst, nor AXP 1E 1048.1$-$5937 \citep{Dib:2009}, as the timing solution presented was not found using a simple polynomial fit, due to instability of the spin-down. As we are primarily concerned with the lower limits of timing noise, we also ignore the timing properties of SGRs, which are only timed in a phase connected fashion for short periods following outbursts where timing noise would likely be elevated by the burst activity. In quiescence SGRs are faint and have not been observed with sufficient regularity for a phase connected solution to emerge. 

While there is no obvious correlation between $B_o$ and the random walk strengths for phase noise ($S_{\rm PN}$) or spin-down noise ($S_{\rm SN}/\dot{\Omega}^2$) in Figure \ref{strengthplot}, there appears to be a weak correlation of the FN strength ($S_{\rm FN}/\Omega^2$) with $B_o$ across the range of magnetic field\footnote{The weak correlation of this timing noise with magnetic field for lower field strengths was previously noted, see e.g. Figure 9 of \citet{Hobbs:2010}, where timing noise is measured by $\sigma_z(10 {\rm yr})$. The sharp increase in $\sigma_z(10{\rm yr})$ above $B_o \sim 10^{12.5} G$ can also be seen in this figure, but is not mentioned by \citep{Hobbs:2010}. In Figure \ref{strengthplot} the trend is clearer due to the inclusion of the AXPs.}. We note, however, that above a field strength of $B_o \go 10^{12.5}$ G, the lower bound of the FN strength distribution rises sharply with $(S_{\rm FN}/\Omega^2)_{\rm min} \sim \dot{\Omega}^2/\Omega^6 \sim B_o^4$. We focus on FN as the change in $S_{\rm FN}$ along this lower envelope is larger than the intrinsic scatter of the distribution for FN (particularly due to the inclusion of AXP timing data). Thus, we infer that timing noise has a component which depends strongly on the magnetic field and becomes dominant above $B_o \go 10^{12.5}$ G such that $(\sigma_{\rm TN})_{\rm min} \sim B_o^2 \Omega T^{3/2}$. These scalings are within the $2\sigma$ confidence intervals for the timing noise scalings inferred by \citep{Shannon:2010} for magnetars, except for the dependence on $T$, which may be different due to their inclusion of SGRs after outburst to evaluate the mean timing properties of the population. 

Caution must be used in interpreting scaling laws with inferred $B_0^2 \sim
\dot\Omega/\Omega^3$ because any correlation is based on the same
dynamical quantities. With this in mind, we consider two different physical models of timing noise due to magnetospheric variability, torque variation, and moment of inertia variation.

\section{Magnetospheric Torque Variation}
Spin-down torque variation and mode-changing are associated with perturbations of the open field lines \citep{Kramer:2006, Lyne:2010}.  The frequency noise strength for torque variability should scale as $S_{\rm FN}/\Omega^2 \sim \dot{\Omega}^2/\Omega^3 \sim B_o^4 \Omega^3$ \citep{Cheng:1987}, which is inconsistent with the scaling discussed above. Thus, while this source of noise may dominate in some pulsars, particularly those where significant pulse-shape changes are observed, it is not the source of the high-$B_o$ timing noise floor evident in Figure \ref{strengthplot}.

\section{Magnetospheric Moment of Inertia}
The angular momentum content of a small volume of the magnetosphere in the inertial frame  \citep{Michel:1973} is $d{\bf L} =({\bf r} \times {\bf S}/c^2) dV$  
where ${\bf S}$ is the Poynting vector and $dV$ is the volume element. In a co-rotating ideal-MHD magnetosphere with angular velocity ${\bf \Omega}$, we can evaluate this as

\be
d{\bf L} =\left[{\bf r} \times  ({\bf \Omega} \times {\bf r}) \frac{B^2}{4\pi  c^2} - ({\bf r} \times {\bf B}) \frac{{\bf B} \cdot  ({\bf \Omega} \times {\bf r})}{4\pi  c^2}\right]dV \label{Leq}.
\ee

For a dipole with magnetic moment ${\bm \mu}_o$, where $|{\bm \mu}_o| = B_o r_{\rm NS}^3$ and $2B_o$ is the field strength at the magnetic pole at the NS surface $r_{\rm NS}$,  we calculate the magnetospheric angular momentum (for light-cylinder radius $\varpi_{LC}=c/\Omega \gg r_{\rm NS}$) using \eqref{Leq}
\be
{\bf L}_B \simeq \frac{16}{15}\frac{{\mu}_o^2{\bf \Omega}}{c^2 r_{\rm NS}} 
+ \frac{1}{15 c^2 r_{\rm NS}} (\mu_o^2 {\bf \Omega} - [{\bm  \mu_o} \cdot {\bf \Omega}]{\bm \mu}_o).
\ee

For the aligned rotator  this gives ${\bf L}_{\rm B, aligned} \simeq (16/15) B_o^2 r_{\rm NS}^5 {\bf \Omega}/c^2$,
while for the oblique rotator we have ${\bf L}_{\rm B, oblique} \simeq (17/15) B_o^2 r_{\rm NS}^5 {\bf \Omega}/c^2$.

For a magnetic field strength $B_{15} \equiv B_o/10^{15}$ G, NS radius $r_{6} \equiv r_{\rm NS}/(10^6\, {\rm cm})$ and mass $M_{1.4} = M_{\rm NS}/(1.4 M_{\odot})$ we can compare the moment of inertia in the magnetosphere to that in the NS,
$I_{\rm B}/I_{\rm NS} \simeq 10^{-6} B_{15}^2 r_{6}^3 M_{1.4}^{-1} \left(\eta/0.4\right)^{-1}$,
for the aligned rotator, where $\eta \equiv I_{\rm NS}/(M_{\rm NS}r_{\rm NS}^2) = 2/5$ for a uniform rotating sphere. While this ratio depends strongly on the NS radius, typical equations of state which allow masses as large as the observed $2M_\odot$, have radii varying by at most $\sim 20\%$ over the range of expected NS masses \citep{Demorest:2010, Steiner:2013}.

Magnetospheric plasma has at least the Goldreich-Julian (GJ) density $\rho_{\rm GJ}={\bf \Omega}\cdot {\bf B}/[2 \pi c(1-(\varpi/\varpi_{\rm LC})^2]$ \citep{GoldreichJulian:1969}, where $\varpi$ is the cylindrical radius. Near the NS surface, $\varpi\ll \varpi_{LC}$, the ratio of the GJ plasma energy density to the magnetic field energy density is $E_{\rm GJ}/E_{\rm mag}\sim10^{-19} (\Omega/s^{-1}) (\mu_{o}/10^{30}{\rm G~cm^3})^{-1} (\varpi/10^{6}{\rm cm})^{3}$. The radius $\varpi_{\rm eq}$ at which the energy density of the plasma is comparable with the magnetic field energy density can be estimated as $\varpi_{\rm eq}=\varpi_{\rm LC}\sqrt{1- 4 c^{4}m_{e}/(\mu_{o}\Omega^{2}{\rm e})}$, i.e. $\varpi_{\rm eq} \simeq 0.999\varpi_{\rm LC}$ for $B_o \simeq 10^{12}$G and $\Omega \simeq 1 s^{-1}$.  In this calculation we have assumed corotation of the magnetospheric plasma. Allowing more complicated motion does not change this result but it can decrease the radius where the two energy densities are equal. Relativistic force-free solutions of the magnetosphere that take into account the poloidal currents \citep{Contopoulos:1999} lead to similar results as their differences are concentrated near the light cylinder, while most of the mass and energy density is near the NS surface. Even for plasma density several orders of magnitude larger than for GJ \citep{Rafikov:2005} the contribution of the plasma to the moment of inertia of the magnetosphere can be safely ignored.

\begin{figure}
\includegraphics[width=\columnwidth]{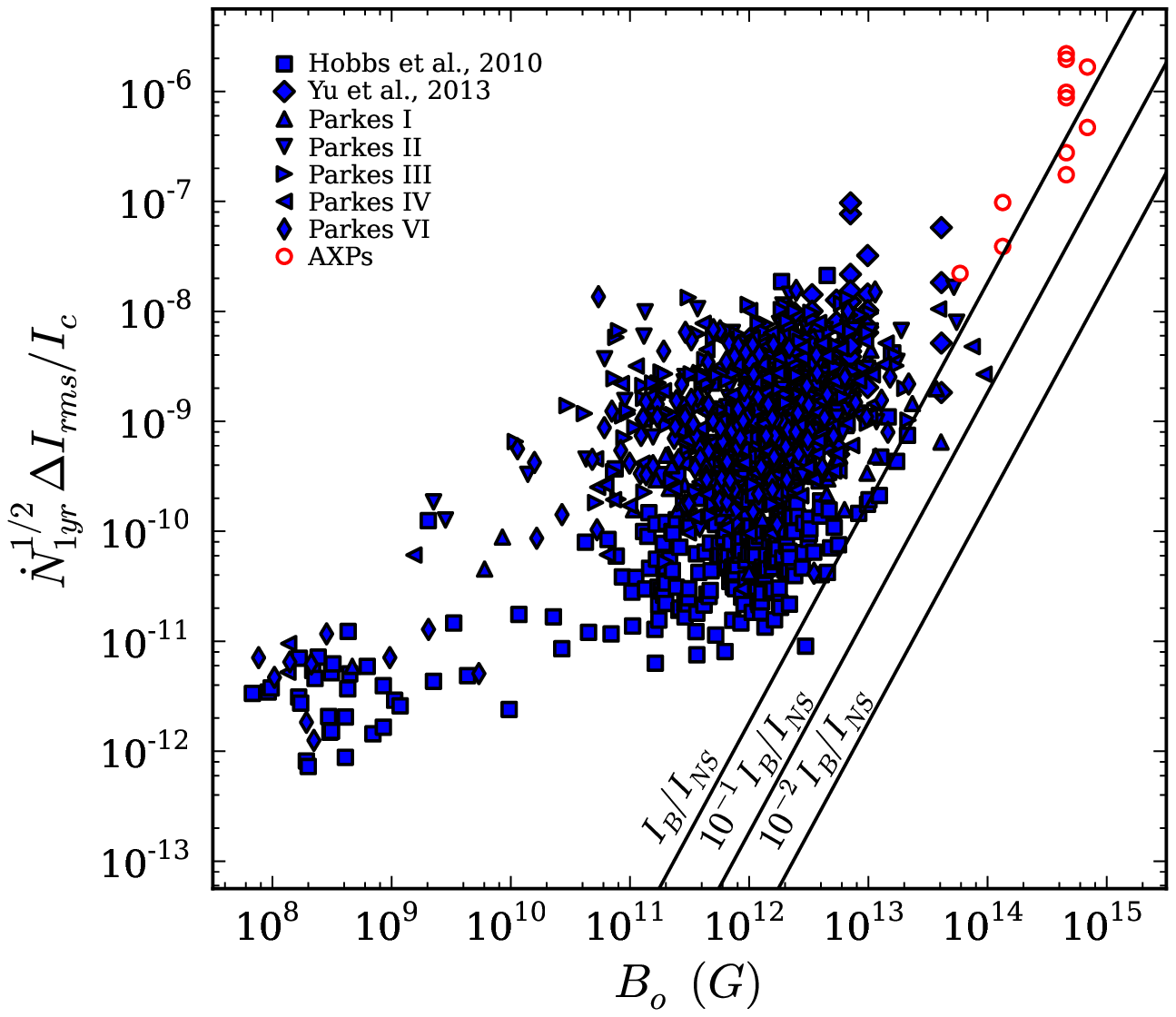}
\caption{A measure of the moment of inertia variability $\dot{N}_{1{\rm yr}}^{1/2} \Delta I_{\rm rms}/I_{c}$ versus $B_o$ for the same pulsars and magnetars shown in Figure \ref{strengthplot}. $\dot{N}_{1 {\rm yr}}$ is the number of moment of inertia random walk steps that occur per year. We compare this to the fractional moment of inertia contained in the magnetosphere, $I_{\rm B}/I_{\rm NS}$ (assuming an aligned dipole for a $1.4 M_\odot$ mass NS with radius 12 km).\label{noiseplot} }
\end{figure}

\section{Magnetospheric Variability}
Mode-changing and nulling events are related to rapid variability of the open field line region of the magnetosphere \citep{Kramer:2006, Lyne:2010}. Recent observations have shown that such rapid variability may be a global magnetospheric phenomenon \citep{Hermsen:2013}, and not simply confined to the open field lines. Thus, the magnetospheric moment of inertia could also vary on a short timescale.

The equation of motion for rotation is $\frac{d}{dt}[I(t) \Omega(t)] = N(t)$,
where $N(t)$ is the external torque. We define $I = I_c + \delta I(t)$ where $\delta I(t)$ is a stochastic component to the moment of inertia associated with magnetospheric variations, and $I_c$ is the moment of inertia of the NS that is strongly coupled to the crust such that the coupling timescale is much shorter than the timescale of the variability. This component is the part of the star that can then respond effectively to the moment of inertia variation. We also define $\Omega(t) = \Omega_S(t) + \delta \Omega(t)$ where $\Omega_S(t)$ is the smooth polynomial angular velocity and $\delta \Omega(t)$ is the stochastic component. Ignoring the spin-down torque variations discussed above, we solve for the $\delta \Omega(t)$ to first order in $\delta I/I_c$ and $\delta \Omega/\Omega_S$, assuming that the torque takes the form $N = \alpha \Omega^n$ where $n$ is the braking index of the NS. This gives
\be
\delta \Omega (t) = -\Omega_S (t) \frac{\delta I(t)}{I_c} - \int^t n\dot{\Omega}_S(t') \frac{\delta I(t')}{I_c} dt' \label{deltaOmegaeqn}
\ee
The second term is much smaller than the first term if the observed time span is $T \ll \tau_c \equiv \Omega_S/2\dot{\Omega}_S \go 10^4$ years for typical magnetars.

Modeling the stochastic variation of the moment of inertia as a random walk we have $\delta I(t) = \sum_j \Delta I_j H(t-t_j)$,
where $\Delta I_j$ and $t_j$  are random amplitudes and times, while $H(t)$ is the unit step function. 
The first term of equation \eqref{deltaOmegaeqn} then corresponds to FN, while the second term is SN. Considering realistic observing spans the first term must dominate. 
Using the definition of $S_{\rm FN}$, the variability of the moment of inertia can then be expressed as
\be
\dot{N}_{1 {\rm yr}}^{1/2} \Delta I_{\rm rms}/I_{c} = \sqrt{(S_{\rm FN}/\Omega^2) \times 1\, {\rm yr}}\,,
\ee
where $\dot{N}_{1{\rm yr}} \equiv R \times 1\, {\rm yr}$ is the number of random walk steps per year. 
This value is plotted against $B_o$ in Figure \ref{noiseplot}, as well as various fractions of the magnetospheric moment of inertia $I_B/I_{\rm NS}$, assuming an aligned magnetic dipole and a NS mass $1.4 M_\odot$ and radius $12$ km. We note here that above $B_o \simeq 10^{12.5}G$ the moment of inertia variability is bounded by $\dot{N}_{1 {\rm yr}}^{1/2} \Delta I_{\rm rms}/I_c \sim (0.1 -1) I_{\rm B}/I_{\rm NS}$. If the timing noise observed at this lower bound is due to variability of the moment of inertia it follows then that
$\dot{N}_{1 \rm yr}^{1/2} (\Delta I_{\rm rms}/I_{\rm B}) (I_{\rm NS}/I_{c}) \sim 0.1 - 1$.

Assuming $\dot{N}_{1 {\rm yr}}\sim 10^5 - 10^7$ to reflect the variability timescales observed in pulsar nulling or mode-changing \citep{Lyne:2010}, and the strongly coupled fraction of the NS moment of inertia to be $I_c/I_{\rm NS} \sim 0.1$\citep{Cheng:1987} over the variability timescale, we can estimate the rms amplitude of the magnetospheric moment of inertia variability, 
$\Delta I_{\rm rms} \sim (10^{-4} - 10^{-6})\, I_{\rm B}.$ Magnetars that are visible in radio vary in timescales ranging from minutes to days depending on the radio frequency \citep{Camilo:2006}, hinting at varying $\dot{N}_{1 {\rm yr}}$ with radius, yet still allowing reasonably small fluctuations of the magnetospheric moment of inertia.

The magnetosphere is expected to be dynamic near the light cylinder, due to reconnection and instability \citep{Contopoulos:1999, Spitkovsky:2006}. 
However, the amplitude of this contribution to the moment of inertia variation is suppressed by a factor $(r_{\rm NS}\Omega/c)^2 \sim 10^{-9}{\rm  s}^2\, \Omega^2$ as the magnetic field is much weaker there. This implies that the variability of the moment of inertia must instead occur near the NS surface, where the magnetospheric moment of inertia is largest.

\section{Comparison to Glitch Sizes}
Glitches in pulsars and magnetars are impulsive increases in the rotation frequency. The smallest detectable glitch size can be estimated from the timing noise level, by comparing the (pre-fit) phase change due to the stochastic noise over the data span required to infer the existence of small glitch, to the phase change due to the glitch itself, $\Delta \phi_{\rm TN}(\Delta t) \simeq S_{\rm FN}^{1/2} (\Delta t)^{3/2}/\sqrt{12} \lo \Delta \Omega_{\rm glitch} \Delta t$ for frequency noise. If the expected phase change due to the noise is larger than the phase change due to the glitch, then a glitch cannot be definitively identified. 
In the continuous limit where the details of cadence and fitting can be simplified we can then estimate the smallest observable glitch size by assuming that several ($\sim 3$) time-of-arrival observations, with cadence $\sim 1$ month, on either side of a small glitch are needed to characterize it. This gives the estimate $(\Delta \Omega/\Omega)_{\rm glitch} \go(0.02 - 0.2) I_{\rm B}/I_{\rm NS}$.

\begin{figure}
\includegraphics[width=\columnwidth]{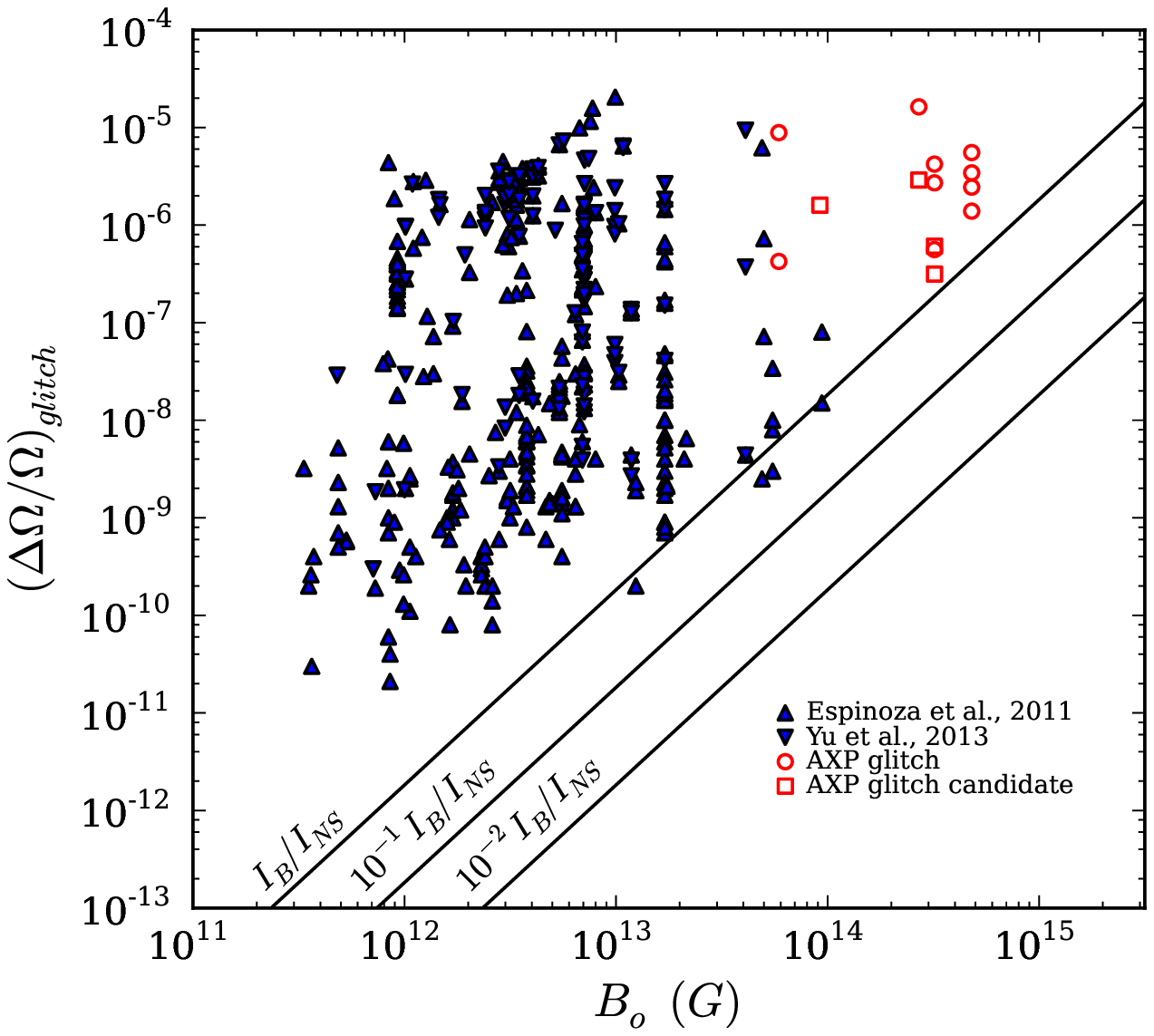}
\caption{Relative glitch size, $\Delta \Omega/\Omega$, versus $B_o$, for radio pulsar glitches \citep{Espinoza:2011, Yu:2013}, and for AXP glitches \citep{Gavriil:2002, Dib:2008, Dib:2009}. The ratio of the magnetospheric moment of inertia to the total moment of inertia ($I_{\rm B}/I_{\rm NS}$), is also plotted as a function of $B_o$, along with 10\% and 1\% of $I_{\rm B}/I_{\rm NS}$. The minimum observed glitch size serves as a proxy for the timing noise, scaling with $(\Delta \Omega/\Omega)_{\rm min}  \sim 0.3 I_{\rm B}/I_{\rm NS}$ for $B_o \go 10^{13}G$ and is consistent with our estimates of timing noise.\label{glitch plot}}
\end{figure}

In Figure \ref{glitch plot} we show the relative glitch sizes $(\Delta \Omega/\Omega)_{\rm glitch}$ as a function of magnetic field for the glitching pulsars listed in the literature \citep{Espinoza:2011, Yu:2013}. We also include glitches and glitch candidates from AXPs 4U 0142$+$61 \citep{Dib:2007, Gavriil2011}, 1E 2259.1$+$586 \citep{Kaspi:2003, Dib:2008}, 1E 1841$-$045 \citep{Dib:2008}, RXS J170849.0$-$400910 \citep{Dib:2008} and 1E 1048.1$-$5937 \citep{Dib:2009}. 
We find that for $B_o \go 10^{13}$ G, the minimum observed glitch is roughly given by $(\Delta \Omega/\Omega)_{\rm glitch} \go 0.3 I_{\rm B}/I_{\rm NS}$, which is consistent with our estimates above. Thus, we find consistent evidence from glitches for an increase with $B_o$ of the timing noise floor. 

While in principle a glitch due to a change in the magnetospheric moment of inertia could be detected, it would require a large ($\go 10\%$) change in the total magnetospheric moment of inertia to be above the timing noise. Such an event would almost certainly be accompanied by torque variations and particle outflows, as seen during giant flares in SGRs, which would dominate the timing change due to the magnetospheric moment of inertia.

\section{Discussion}
We have examined pulsar and AXP timing noise measurements reported in the literature and shown that a component of timing noise exists which depends on the spin-inferred dipole surface magnetic field strength ($S_{\rm FN}/\Omega^2 \sim B_o^4$ such that $\sigma_{\rm TN} \sim B_o^2 \Omega T^{3/2}$). This timing noise component begins to dominate at $B_o \sim 10^{12.5}$ G, and is responsible for a sharp rise in the floor of the timing noise values across both pulsar and AXP populations. 

This provides yet another connection between high-B radio pulsars and AXPs, demonstrating a continuum of behaviors in these NSs, as independently suggested by, for example,  quiescent X-ray luminosities of these objects \citep{An:2012}, further `unifying' radio pulsars and AXPs \citep{Kaspi:2010}. 

Variations near the open field lines can lead to mode changing and torque variability \citep{Lyne:2010}, however this would result in a frequency noise strength that scales as $S_{\rm FN}/\Omega^2 \sim B_o^4 \Omega^{3}$, and cannot explain the observed timing noise dependence. 

We have shown that the magnetospheric moment of inertia is $I_{\rm B} \simeq 10^{-6} B_{15}^2 I_{\rm NS}$, and have proposed a model of magnetospheric moment of inertia variation that is consistent with the observations of both the timing noise strengths and the size of the smallest observable glitches in high-$B_o$ systems. 
By assuming a rate similar to the known variability in the open field line regions of some pulsars \citep{Kramer:2006, Lyne:2010} we can estimate an amplitude for the variability of the magnetospheric moment of inertia $\Delta I_{\rm rms} \sim (10^{-4} - 10^{-6}) I_{\rm B}$. 
This variation must occur near the NS surface, where the moment of inertia contribution is largest, as we find that timing noise due to variability confined near the light cylinder is too small to contribute significantly. 

Recent observations \citep{Hermsen:2013} have also provided evidence that there exists rapid global variability in pulsar magnetospheres. 
We suggest that rapid global magnetospheric variability, perhaps due to reconnection or variable currents which have been proposed in pulsar models
\citep{Contopoulos:2005, Li:2012}, acts as a source of timing noise through moment of inertia variations.

Other potential sources of this magnetic field-timing noise dependence that could also be considered are the internal field evolution as a source of moment of inertia variability, or  the interaction of the superfluid vortices with the magnetic field affecting glitch dynamics. 


{\it Acknowledgments}--- 
DT was supported by funding from the Lorne Trottier Chair in Astrophysics and Cosmology, and the Canadian Institute for Advanced Research. KNG was supported by the Centre de Recherche en Astrophysique du Qu{\'e}bec. We would like to thank Vicky Kaspi, Andrew Cumming, Anne Archibald, Rob Archibald, Jim Cordes, Joanna Rankin, Ioannis Contopoulos, Maxim Lyutikov, Chris Hirata and Peter Goldreich for insightful discussions during the course of this work.


\end{document}